\documentclass[aps,prd,preprintnumbers,nofootinbib]{revtex4}
\usepackage[dvipdfmx]{graphicx}
\usepackage{bm,latexsym,amsmath,amsthm,amssymb,amsfonts,color}

\usepackage[normalem]{ulem}

\begin{document} 
\title{\uline{}On uniqueness of static black hole with conformal scalar hair}

\author{${}^{1}$Yoshimune Tomikawa, ${}^{2,3}$Tetsuya Shiromizu and ${}^{3,2}$Keisuke Izumi}
\affiliation{${}^1$School of Informatics and Sciences, Nagoya University, Nagoya 464-8601, Japan}
\affiliation{${}^2$Department of Mathematics, Nagoya University, Nagoya 464-8602, Japan}
\affiliation{${}^3$Kobayashi-Maskawa Institute, Nagoya University, Nagoya 464-8602, Japan}
\begin{abstract}
We discuss the uniqueness of the static black hole in the Einstein gravity with a conformally coupled 
scalar field. In particular, we prove the uniqueness of the region outside of the photon surface, 
not event horizon. 
\end{abstract}

\maketitle

\section{Introduction} \label{sec-intro}

It is well-known that static black holes do not have the scalar hair with non-negative 
potential in asymptotically flat spacetimes \cite{bekenstein1972} (see Ref. \cite{H-R} for a review). 
However, there is an exception, that is, we can have the Bocharova-Bronnikov-Melnikov-Bekenstein (BBMB) 
solution \cite{B-B-M, bekenstein1974}. This is because this solution does not satisfy 
the regularity condition for the scalar field at the event horizon.

In this paper, we shall address the uniqueness issue of the BBMB black hole solution. 
In particular, we will examine if the static black hole spacetime with conformal scalar 
hair is spherically symmetric as the Schwarzschild spacetime \cite{Israel}. As a result, 
we could prove the uniqueness of the photon surface of the BBMB solution, that is, the 
outside region of the photon surface is unique to be the BBMB solution in the Einstein 
gravity with a conformally coupled scalar field. In the BBMB solution, the photon surface 
corresponds to the unstable circular orbit of null geodesics (see Ref. \cite{Claudel} 
for the definition of the photon surface). In the Einstein frame, the system is reduced to 
the Einstein-massless scalar field system. In this system, the uniqueness of the outiside 
region of the photon surface has been proven in Ref. \cite{Y}. However, the proof in Ref. 
\cite{Y} cannot be applied to the current situation. This is because the photon surface 
of the BBMB black hole in the Jordan frame is singular in the Einstein frame. 

Note that it will be difficult to prove the black hole uniqueness for cases except for 
vacuum/electrovacuum or well-motivated systems by string theory or so \cite{Heusler}. 
In this sense, the current result may encourage us to try to prove 
the uniqueness of static black holes with a hair although we know that the BBMB black 
hole itself is not stable \cite{H-R}. 

The rest of this paper is organized as follows. In Sec. 2, we briefly review the BBMB 
black hole. In Sec. 3, we describe the basic equations for static spacetimes in the 
Einstein gravity with the conformally coupled scalar field. In Sec. 4, we show that 
the certain relation between the scalar field and the time lapse function holds once 
the scalar field is turned on. Then, in Sec. 5, we will present the proof of the 
uniqueness of the photon surface in the current system. Finally, we will give the summary 
and discussion in Sec. 6. 

\section{BBMB black hole}

Let us consider the Einstein equation with the conformally coupled scalar field 
\cite{B-B-M, bekenstein1974},
\begin{eqnarray}
S=\dfrac{1}{2\kappa} \displaystyle \int d^4 x\sqrt{-g} R-\int d^4 x\sqrt{-g} \Big( \dfrac{1}{2} (\nabla \phi)^2 
+\dfrac{1}{12} R\phi ^2 \Big), \label{BBMBaction}
\end{eqnarray}
where $\phi$ is the scalar field and $R$ is the Ricci scalar. The field equations are 
\begin{eqnarray}
G_{\mu \nu}=\kappa T_{\mu \nu} \label{BBMB-Eineq} 
\end{eqnarray}
and
\begin{eqnarray}
\nabla^2 \phi=\dfrac{1}{6}R\phi, \label{BBMB-scalareq}
\end{eqnarray}
where $G_{\mu \nu}$ is the Einstein tensor and
\begin{eqnarray}
T_{\mu \nu}= \nabla_\mu \phi \nabla_\nu \phi -\dfrac{1}{2} g_{\mu \nu} (\nabla \phi)^2 +\dfrac{1}{6} (g_{\mu \nu} \nabla ^2 -\nabla_\mu \nabla_\nu +G_{\mu \nu}) \phi^2. \label{T}
\end{eqnarray}
The trace for Eq. (\ref{BBMB-Eineq}) and Eq. (\ref{BBMB-scalareq}) show us
\begin{eqnarray}
R=0,\label{rs0}
\end{eqnarray}
and 
\begin{eqnarray}
\nabla ^2 \phi =0. \label{BBMB-phi=0}
\end{eqnarray}
For the current purpose, it is better to rearrange the Einstein equation as 
\begin{eqnarray}
\Bigl(1-\frac{\kappa}{6}\phi^2\Bigr)R_{\mu\nu}=\kappa S_{\mu\nu},\label{einstein2}
\end{eqnarray}
where 
\begin{eqnarray}
S_{\mu\nu}:=\frac{2}{3}\nabla_\mu \phi \nabla_\nu \phi-\frac{1}{6}
g_{\mu\nu}(\nabla \phi)^2-\frac{1}{3}\phi \nabla_\mu \nabla_\nu \phi. \label{smunu}
\end{eqnarray}
From this, we can see that one needs a careful treatment at $\phi=\pm {\sqrt {6/\kappa}}=:\phi_p$ 
to have the regular spacetime. 

The metric of the BBMB black hole is given by \cite{B-B-M, bekenstein1974}
\begin{eqnarray}
ds^2=-f(r) dt^2 + f^{-1}(r) dr^2 +r^2 d\Omega_2^2,
\end{eqnarray}
where $f(r)=(1-m/r)^2$, $m$ is the mass of black hole and $d\Omega_2^2$ is 
the metric of the unit 2-sphere. The configuration of the scalar field is 
\begin{eqnarray}
\phi =\pm \sqrt{\dfrac{6}{\kappa}} \dfrac{m}{r-m}.
\end{eqnarray}
The metric itself is exactly the same with the extreme Reissner-Nordstr\"{o}m black hole 
spacetime. The event horizon is located at $r=m$ and the scalar field diverges at there. 
Here we note that the factor $1-\kappa \phi^2/6$ in the left-hand side of Eq. 
(\ref{einstein2}) vanishes at $r=2m$ where there is the unstable circular orbit of null 
geodesics.

\section{Basic equations in static spacetimes}

Let us focus on static spacetimes from now on. Then the metric is written as 
\begin{eqnarray}
ds^2=-V^2(x^k)dt^2 +g_{ij}(x^k) dx^i dx^j.
\end{eqnarray}
The Latin indices indicate the spatial components. The locus of the event horizon is 
$V=0$ and we assume that the spacetime is regular in the outside of the event horizon, that is, 
the domain of the outer communication. 

In static spacetimes, the non-trivial parts of the Ricci tensor are 
\begin{eqnarray}
R_{00}=VD^2 V \label{ricci00}
\end{eqnarray}
and 
\begin{eqnarray}
R_{ij} ={}^{(3)} R_{ij} -V^{-1} D_i D_j V, \label{static-Rij}
\end{eqnarray}
where $D_i$ and ${}^{(3)}R_{ij}$ are the covariant derivative and the Ricci tensor 
on the $t=$constant hypersurface $\Sigma$, respectively. 
$S_{\mu\nu}$ defined in Eq. (\ref{smunu}) is decomposed into 
\begin{eqnarray}
& & S_{00}=\frac{1}{6}[V^2(D\phi)^2+2\phi V D^i V D_i \phi], \\
& & S_{ij}=\frac{2}{3}D_i \phi D_j \phi -\frac{1}{6}g_{ij}(D\phi)^2
-\frac{1}{3}\phi D_i D_j \phi.
\end{eqnarray}
The equation for the scalar field (\ref{BBMB-phi=0}) becomes 
\begin{eqnarray}
D_i (VD^i \phi)=0. \label{dvdp}
\end{eqnarray}

Since we focus on asymptotically flat cases, the asymptotic behaviors of metric 
are given by 
\begin{eqnarray}
V=1-m/r+O(1/r^2)
\end{eqnarray}
and
\begin{eqnarray}
g_{ij}=(1+2m/r)\delta_{ij}+O(1/r^2). 
\end{eqnarray}
For the scalar field, we impose
\begin{eqnarray}
\phi=O(1/r). 
\end{eqnarray}

From Eqs. (\ref{rs0}), (\ref{ricci00}) and (\ref{static-Rij}), we can see   
\begin{eqnarray}
{}^{(3)}R=2V^{-1} D^2 V. \label{static-R=0}
\end{eqnarray}

\section{Scalar field and time lapse function}

In this section, we will show that the scalar field $\phi$ is written by the time lapse function $V$ uniquely 
if there is non-trivial scalar field exists in static spacetimes. 

From the $(0,0)$-component of the Einstein equation and Eq. (\ref{dvdp}), we have 
\begin{eqnarray}
D_i [(1-\varphi)D^i \Phi]=0,\label{new}
\end{eqnarray}
where $\Phi:=(1+\varphi)V$ and $\varphi:=\pm {\sqrt {\kappa /6} \phi}$. Now, in $\Sigma$, 
we focus on 
the region $\Omega$ which has the two boundaries; the surface $S_p$ specified by $\phi=\phi_p$ 
and the 2-sphere $S_\infty$ at the spatial infinity. In $\Omega$, we assume that there are 
no event horizons, that is, $V$ is strictly positive ($V>0$). 
Since $\varphi$ or $\phi$ follows Eq. (\ref{dvdp}), 
$\varphi$ is a monotonic function which has the maximum value $1$ at 
$S_p$ and minimum value $0$ at the spatial infinity, 
that is, $0 \leq \varphi \leq 1$ in $\Omega$. 

Let us consider the conformally transformed space $\tilde \Omega (\subset \tilde \Sigma)$ with the 
metric $\tilde g_{ij}=(1-\varphi)^2 g_{ij}$. Then Eq. (\ref{new}) becomes 
\begin{eqnarray}
\tilde D^2 \Phi=0.
\end{eqnarray}
The volume integration of the above over $\tilde \Omega$ and 
the Gauss theorem give us 
\begin{eqnarray}
0=\int_{\tilde \Omega} \tilde D^2 \Phi d \tilde \Sigma =\int_{\tilde S_\infty} \tilde D_i \Phi d \tilde S^i
-\int_{\tilde S_p} \tilde{D}_i \Phi d \tilde S^i.\label{int1}
\end{eqnarray}
We can show that the second term of the right-hand side vanishes as 
\begin{eqnarray}
\int_{\tilde S_p}\tilde D_i \Phi d \tilde S^i=\int_{S_p}(1-\varphi)D_i \Phi dS^i=0. \label{sidpps}
\end{eqnarray}
For the last equality in the above, we used the fact of $\varphi|_{S_p}=1$. 
Then, Eq. (\ref{int1}) tells us 
\begin{eqnarray}
\int_{\tilde S_\infty}\tilde D_i \Phi d \tilde S^i=0. \label{sidpsi}
\end{eqnarray}
Next we consider the volume integration of $\Phi \tilde D^2 \Phi=0$ over $\tilde \Omega$ and then 
\begin{eqnarray}
0& = & \int_{\tilde \Omega} \Phi\tilde D^2 \Phi d \tilde \Sigma \nonumber \\
& = & -\int_{\tilde \Omega} (\tilde D \Phi)^2 d \tilde \Sigma
+\int_{\tilde S_\infty}\Phi\tilde D_i \Phi d \tilde S^i
-\int_{\tilde S_p}\Phi \tilde D_i \Phi d \tilde S^i. \nonumber \\
& & \label{sipdp}
\end{eqnarray}
In the second equality, we used the Gauss theorem. As Eq. (\ref{sidpps}), it is easy to show that 
the last term vanishes. Using Eq. (\ref{sidpsi}) and the fact of $\Phi_\infty=1$, 
we can see 
\begin{eqnarray}
\int_{\tilde S_\infty}\Phi\tilde D_i \Phi d \tilde S^i=\int_{\tilde S_\infty}\tilde D_i \Phi d \tilde S^i=0.
\end{eqnarray}
Thus, Eq. (\ref{sipdp}) implies 
\begin{eqnarray}
\int_{\tilde \Omega} (\tilde D \Phi)^2 d \tilde \Sigma=\int_{\Omega} (D \Phi)^2(1-\varphi) d \Sigma= 0.
\end{eqnarray}
Therefore, $D_i \Phi=0$ has to be satisfied everywhere on $\Omega$. Together with the boundary condition at the spatial infinity ($\Phi_\infty=1$), 
this means that $\Phi=1$ holds everywhere. Thus we have 
the following relation between the time lapse function $V$ and 
the scalar field $\phi$;
\begin{eqnarray}
\phi=\pm \sqrt{\dfrac{6}{\kappa}} (V^{-1}-1). \label{pvrel}
\end{eqnarray}
Of course, the BBMB solution satisfies this relation. Note that $\phi=\phi_p$ 
corresponds to $V=1/2=:V_p$ and the $V=V_p$ surface in the BBMB solution is 
composed of the closed circular orbit of photon (null geodesics). 

Now, both Eq. (\ref{dvdp}) and the $(0,0)$-component of the Einstein equation 
give us the same equation 
\begin{eqnarray}
D^2 v = 0, \label{laplace}
\end{eqnarray}
where $v:=\ln V$. Then, we can use $v$ as a ``radial" coordinate later. 

The $(i,j)$-component of the Einstein equation becomes 
\begin{eqnarray}
\frac{2V-1}{V^2}\Bigl({}^{(3)}R_{ij}-\frac{1}{V}D_iD_jV \Bigr) 
=4\frac{D_iVD_jV}{V^4}-g_{ij}\frac{(DV)^2}{V^4}-2\Bigl(\frac{1}{V}-1 \Bigr)D_i D_j V^{-1}. \label{eij}
\end{eqnarray}
The trace part of the above and Eq. (\ref{laplace}) (or Eqs. (\ref{static-R=0}) 
and (\ref{laplace})) imply 
\begin{eqnarray}
{}^{(3)}R=\frac{2}{V^2}(DV)^2 \geq 0. \label{3riccidv2}
\end{eqnarray}

\section{Uniqueness of the BBMB photon surface}

In this section, we will employ Israel's way to 
address the uniqueness of the BBMB black hole.

First, we consider the foliation by $v=$constant 2-surfaces $\lbrace S_v \rbrace$ 
on $t=$constant hypersurfaces $\Sigma$. We write the induced metric on 2-surfaces as $h_{ij}:
=g_{ij}-n_i n_j$, where $n_i:=\rho D_i v$ is the unit normal vector on the 
surface and $\rho :=(D_i v D^i v)^{-1/2}=V(D_iVD^iV)^{-1/2}$. Then we can derive 
the following equations;
\begin{eqnarray}
D_i \Bigl( \frac{n^i}{\rho} \Bigr) =0, \label{div1}
\end{eqnarray}
\begin{eqnarray}
D_i \Bigl(\frac{(\rho k-2)n^i}{(2V-1)\rho^{3/2}}  \Bigr) 
=-\frac{1}{2V-1}(\rho^{-1/2}\tilde k_{ij}\tilde k^{ij}+\rho^{-3/2}{\cal D}^2\rho )
\label{div2}
\end{eqnarray}
and
\begin{eqnarray}
D_i \Bigl( (k\xi+\eta)n^i \Bigr)=-(\tilde k_{ij}\tilde k^{ij}+\rho^{-1}{\cal D}^2 \rho )\xi,\label{div3}
\end{eqnarray}
where $\xi:=(2V-1)\rho^{-1/2}$, $\eta:=2(2V+1)\rho^{-3/2}$, $k_{ij}$ is  the extrinsic curvature of $S_v$ in $\Sigma$, $k$ is its trace part and $\tilde k_{ij}:=k_{ij}-(1/2)kh_{ij}$ is its traceless part. 
${\cal D}_i$ is the covariant derivative with respect to $h_{ij}$. In the above, we used 
the following equations;
\begin{eqnarray}
n^i D_i V=\frac{V}{\rho},
\end{eqnarray}
\begin{eqnarray}
n^i D_i \rho=\rho k
\end{eqnarray}
and
\begin{eqnarray}
n^i D_i k & = & -k_{ij}k^{ij}-\rho^{-1}{\cal D}^2 \rho-{}^{(3)}R_{ij}n^i n^j \nonumber \\
& = & -k_{ij}k^{ij}-\rho^{-1}{\cal D}^2 \rho+\frac{\rho k-4V}{(2V-1)\rho^2}.
\end{eqnarray} 
For the derivation of the second equation, we used Eq. (\ref{laplace}). 
For the third one, we used Eq. (\ref{eij}) and the formula 
\begin{eqnarray}
D_i D_j V  =  \frac{V}{\rho}\Bigl[ k_{ij}-\frac{1}{\rho}(n_i {\cal D}_j \rho+n_j {\cal D}_i \rho) -n_i n_j \Bigl(k-\frac{1}{\rho} \Bigr) \Bigr]. \label{didjv} 
\end{eqnarray}
This comes from the definition of the extrinsic curvature and Eq. (\ref{laplace}). 

Now we compute the curvature invariant $R_{\mu\nu}R^{\mu\nu}$ to 
check the regularity of spacetime. 
It is written as 
\begin{eqnarray}
R_{\mu\nu}R^{\mu\nu} & = & \frac{1}{\rho^4}+\frac{1}{(2V-1)^2\rho^2}
\Bigl[ \Bigl( 2(1-V)k_{ij}-\frac{1}{\rho}h_{ij} \Bigr)^2 \nonumber \\
& & ~~+\Bigl(-2(1-V)k+\frac{1+2V}{\rho} \Bigr)^2 +\frac{8(1-V)^2}{\rho^2}({\cal D}\rho)^2 
 \Bigr].
\end{eqnarray}
In the above, we used the Einstein equation and Eq. (\ref{didjv}). 

As commented below Eq. (\ref{smunu}), 
one needs a careful treatment at the surface of $V=1/2 ~(\phi=\phi_p)$.  
For the regularity of spacetimes at the surface of $V=1/2$, we require the conditions  
\begin{eqnarray}
{\cal D}_i \rho|_{S_p}=0 \label{rc1}
\end{eqnarray}
and
\begin{eqnarray}
k_{ij}|_{S_p}=\frac{1}{\rho_p}h_{ij}|_{S_p}. \label{rc2}
\end{eqnarray}
Note that the index ``$p$" indicates the evaluation at $S_p$, e.g. $\rho_p=\rho|_{S_p}$. 
These features tell us that $S_p$ is totally umbilic and the photon surface defined in 
Ref. \cite{Claudel} \footnote{Although the theory we consider is different from that in 
Ref. \cite{Claudel}, we can show that $S_p$ is indeed the photon surface. In addition, 
we can easily see that the null geodesic initially tangent to $S_p \times {\bf R}_t$ 
remains to be tangent to $S_p \times {\bf R}_t$.} or photon sphere \cite{Cederbaum:2014gva}. 

From Eqs. (\ref{3riccidv2}) and (\ref{didjv}), the Kretschmann invariant can be written as 
\begin{eqnarray}
& & R_{\mu\nu\rho\sigma}R^{\mu\nu\rho\sigma} \nonumber \\
& & ~~=  \frac{4}{V^2}D_i D_j V D^i D^j V+4{}^{(3)}R_{ij} {}^{(3)} R^{ij}-({}^{(3)}R)^2 \nonumber \\
& & ~~=\frac{4}{\rho^2}\Bigl[k_{ij} k^{ij}+\frac{2}{\rho^2}({\cal D}\rho)^2+\Bigl(k-\frac{1}{\rho} \Bigr)^2\Bigr]
\nonumber \\
& &~~~~+\frac{4}{(2V-1)^2\rho^2}\Bigl[ \Bigl(k_{ij}-\frac{1}{\rho}h_{ij} \Bigr)^2 
+\frac{2}{\rho^2}({\cal D}\rho)^2+\Bigl(k-\frac{4V}{\rho} \Bigr)^2
\Bigr]-\frac{4}{\rho^4},
\end{eqnarray}
where we also used
\begin{eqnarray}
R_{0i0j}=VD_iD_j V,
\end{eqnarray}
and
\begin{eqnarray}
{}^{(3)}R_{ij} =  \frac{1}{(2V-1)\rho}\Bigl[ k_{ij}-\frac{1}{\rho}h_{ij}
-\frac{1}{\rho}(n_i {\cal D}_j \rho+n_j {\cal D}_i \rho)  -n_in_j \Bigl(k-\frac{4V}{\rho} \Bigr) \Bigr].\label{3rij}
\end{eqnarray}
Thus, we can see that the Kretschmann invariant is finite 
in $\Omega$ when Eqs. (\ref{rc1}) and (\ref{rc2}) hold on $S_p$. 

From now on, we focus on the region $\Omega$ in $\Sigma$ which has the two 
boundaries, that is, $S_p$ and the spatial infinity $S_\infty$. 

The volume integrations of Eqs. (\ref{div1}), (\ref{div2}) and (\ref{div3}) over $\Omega$ give us 
\begin{eqnarray}
A_p=4 \pi m \rho_p, \label{equal1}
\end{eqnarray}
\begin{eqnarray} 
 8\pi m^{1/2}-\frac{1}{2}\rho_p^{1/2} \int_{S_p}{}^{(2)}RdS  =-\int_{1/2}^1dV\frac{1}{V(2V-1)} 
\int_{S_v}dS \Bigl( \rho^{1/2}\tilde k_{ij}\tilde k^{ij} 
+\frac{1}{2\rho^{3/2}}({\cal D}\rho)^2 \Bigr) \label{equal2}
\end{eqnarray}
and 
\begin{eqnarray}
 8\pi m^{1/2}-\frac{4A_p}{\rho_p^{3/2}}  =-\int_{1/2}^1dV \frac{(2V-1)}{V} 
\int_{S_v}dS \Bigl( \rho^{1/2}\tilde k_{ij}\tilde k^{ij}
+\frac{1}{2\rho^{3/2}}({\cal D}\rho)^2 \Bigr) , \label{equal3}
\end{eqnarray}
respectively. $A_p$ and ${}^{(2)}R$ are the area of $S_p$ and the Ricci scalar of $S_v$, 
respectively. 
Equation (\ref{equal1}) tells us that $m$ is positive. In the left-hand side of Eq. (\ref{equal2}), 
we used the fact of 
\begin{eqnarray}
{}^{(2)}R=\frac{2}{\rho^2}+k^2-k_{ij}k^{ij}+\frac{2(\rho k-4V)}{(2V-1)\rho^2}. \label{2ricci}
\end{eqnarray}
In particular, on $S_p$, we have 
\begin{eqnarray}
{}^{(2)}R_p=\lim_{V \to 1/2}\frac{2(\rho k-2)}{(2V-1)\rho^2}. \label{2riccip}
\end{eqnarray}
Then Eq. (\ref{equal2}) shows us 
\begin{eqnarray}
16\pi m^{1/2} \leq \rho_p^{1/2}\int_{S_p}{}^{(2)}RdS. \label{miegb}
\end{eqnarray}
Because of $m>0$, inequality (\ref{miegb}) and the Gauss-Bonnet theorem tell 
us that the topology of $S_p$ is restricted to $S^2$(that is, $\int_{S_p}{}^{(2)}RdS=8\pi$) and then we have 
the following inequality with the help of Eq. (\ref{equal1}); 
\begin{eqnarray}
A_p \geq 16 \pi m^2. \label{Sp-geq16}
\end{eqnarray}

Using Eq. (\ref{equal1}), Eq. (\ref{equal3}) gives us the inequality 
\begin{eqnarray}
A_p \leq 16\pi m^2. \label{Sp-leq16}
\end{eqnarray}
This may be regarded as a mimic of the Penrose inequality \cite{Penrose:1973um} 
for the photon surface. 

Thus, we can conclude that the equality holds in inequalities (\ref{Sp-geq16}) and (\ref{Sp-leq16}), 
and then 
\begin{eqnarray}
\tilde k_{ij}=0,  \ {\cal D}_i \rho=0. \label{sp0}
\end{eqnarray}
Finally, we use the Codazzi equation, 
\begin{eqnarray}
{\cal D}_i k^i_j-{\cal D}_j k={}^{(3)}R_{ik}n^i h^k_j, \label{cdz}
\end{eqnarray}
to show that $k$ is constant on each $S_v$. From Eq. (\ref{3rij}), we see that 
the right-hand side vanishes. Using Eq. (\ref{sp0}), we can have 
\begin{eqnarray}
{\cal D}_ik=0. 
\end{eqnarray}
Because of Eq. (\ref{2ricci}), we can see that ${}^{(2)}R$ is also constant on each $S_v$. 
In addition, inequality (\ref{miegb}) tells us that 
${}^{(2)}R$ is positive. Thus, we can conclude that the region of $\Omega$ in the spacetime 
is maximally symmetric with the positive curvature and then spherically symmetric. Thus, we could see 
that the spacetime in $\Omega$ is unique to be the BBMB solution because the regular 
spherically symmetric solutions have been shown to be unique in Ref. \cite{X-Z}.

Note that what we could prove is only the uniqueness for outside region of the photon surface and 
we did not discuss the inside of the photon surface.
Therefore, this proof does not mean the uniqueness of the whole region of the BBMB black hole spacetime. 

\section{Summary and discussion}

In this paper, we proved that, in the Einstein gravity with a conformally coupled scalar field, 
if a single closed surface $S_p$ satisfying $(1-\kappa \phi^2/6 )=0$ 
exists and if there is no horizon in the region $\Omega$ whose boundaries are only $S_p$ and the spatial 
infinity, the geometry in $\Omega$ is the same as that outside of the BBMB photon surface. 
Since we employed Israel's way which can work for the proof 
of the uniqueness of single object, we cannot exclude the existence of multi-photon surface system. 
And we did 
not prove the uniqueness of the whole region of the BBMB black hole spacetime. We have no definite answer for 
the inside region of the photon surface. Accidentally, the uniqueness of the photon surfaces 
have been recently discussed for vacuum, electrovacuum and so on 
\cite{Cederbaum:2014gva,Cederbaum:2015aha,Cederbaum:2015fra,Y,Yoshino}. Therein, the existence 
of photon surface or photon sphere was assumed by hand. On the other hand, 
in our current study on the Einstein gravity with a conformally coupled scalar field, the existence of 
the photon surface is automatically required to make the spacetime regular. 

There are many remaining works. If one may examine the possibility of cases having multi-photon 
surface, one must employ the another proof developed by Bunting and Masood-ul-Alam \cite{Bunting}. 
One may be also interested in the uniqueness issue for the inside region of the photon surface, that is, 
the region between the photon surface and the event horizon. Some extension of 
our proof into other non-vacuum cases should be addressed. Finally, we may have the Penrose inequality 
for a kind of the photon surface for {\it dynamical} systems in the Einstein gravity with conformally coupled 
scalar fields. This is because we could have a mimic of the Penrose inequality in our study. 

\begin{acknowledgments}
This work is initiated by the collaboration with Mr. K. Ueda. 
T. S. is supported by Grant-Aid for Scientific Research from Ministry of Education, 
Science, Sports and Culture of Japan (Nos. 25610055 and 16K05344).
\end{acknowledgments}





\end{document}